\documentclass {article}
\usepackage{graphicx}
\usepackage[small]{subfigure,epsfig}

\usepackage {amsmath} \usepackage{amssymb} \usepackage{cite}

\begin{document}

\title{A Note on "Exp-function method for the exact solutions of fifth order KdV equation and modified Burgers equation"}
\author{Nikolay A. Kudryashov\footnote{nakudryashov@mephi.ru}, Dmitry I. Sinelshchikov}

\date{Department of Applied Mathematics, National  Research Nuclear University MEPHI, 31 Kashirskoe
Shosse, 115409 Moscow, Russian Federation}

\maketitle

\begin{abstract}
We discuss the recent paper by Inan and Ugurlu [Inan I.E., Ugurlu Y., Exp-function method for the exact solutions of fifth order KdV equation and modified Burgers equation, Appl. Math. Comp. 217 (2010) 1294 -- 1299]. We demonstrate that all exact solutions of fifth order KdV equation and modified Burgers equation by Inan and Ugurlu are trivial solutions that are reduced to constants. Moreover, we show exact solutions of the fifth -- order equation studied by Inan and Ugurlu cannot be found by the Exp-function method.
\end{abstract}

\emph{Keywords:} Nonlinear evolution equations, exact solutions, fifth order KdV equation, modified
Burgers equation.

PACS 02.30.Jr - Ordinary differential equations
\\

In the recent paper \cite{Ugurlu} Inan and Ugurlu considered the fifth -- order KdV equation
\begin{equation}
u_{t}+u\,u_{x}+u_{xxxxx}=0
\label{eq1}
\end{equation}

In \cite{Ugurlu}Inan and Ugurlu  used the traveling wave solutions $u(x,t) =u(\xi)$, $\xi=k\,x-w\,t$ in Eq.\eqref{eq1} and obtained nonlinear ordinary differential equation in the form
\begin{equation}
w\,u'+k\,u\,u'+k^{5}u^{(5)}=0\,,
\label{eq2}
\end{equation}
where
\begin{equation*}
u^{'}=\frac{du}{d\xi},\quad u^{(5)}=\frac{d^5\,u}{d\xi^5}.
\label{eq2a}
\end{equation*}

After integration Eq. \eqref{eq2} with respect to $\xi$ we can have
\begin{equation}
C+w\,u+\frac{k}{2}\,u^{2}+k^{5}u^{(4)}=0\,.
\label{eq3}
\end{equation}
Inan and Ugurlu  omitted the constant of integration $C$ in \cite{Ugurlu} and
essentially reduced a class for exact solutions of Eq.\eqref{eq1}. It is known that there are some exact solutions of Eq.\eqref{eq1} expressed through rational and elliptic functions \cite{Kudryashov, Kudryashov04, Demina10}.  However taking the Exp-function method into consideration we cannot find these solutions.

Using Exp-function method Inan and Ugurlu  found the following five solutions of Eq.\eqref{eq1} in \cite{Ugurlu}
\begin{equation}
u_1(x,t)=\frac{-\frac{2w}{k} \exp(kx+wt)-\frac{2b_{0}w}{k}-\frac{2b_{-1}w}{k}\exp(-kx-wt)}{\exp(kx+wt)+b_{0}+b_{-1}\exp(-kx-wt)},
\label{eq4a}
\end{equation}
\begin{equation}
u_2(x,t)=\frac{-\frac{2w}{k} \exp(kx+wt)-\frac{2b_{0}w}{k}}{\exp(kx+wt)+b_{0}},
\label{eq5a}
\end{equation}
\begin{equation}
u_3(x,t)=\frac{-2k^{4} \exp(kx+k^{5}t)-2b_{0}k^{4}}{\exp(kx+k^{5}t)+b_{0}},
\label{eq6a}
\end{equation}
\begin{equation}
u_4(x,t)=\frac{-32k^{4} \exp(kx+16k^{5}t)}{\exp(kx+16k^{5}t)},
\label{eq7a}
\end{equation}
\begin{equation}
u_5(x,t)=\frac{-32k^{4} \exp(kx+16k^{5}t)-32b_{0}k^{4}}{\exp(kx+16k^{5}t)+b_{0}}.
\label{eq8a}
\end{equation}

However it is easy to note that all these "solutions" can be written as constants
\begin{equation}
u_1(x,t)=\frac{2w}{k},
\label{eq4}
\end{equation}
\begin{equation}
u_2(x,t)=-\frac{2w}{k},
\label{eq5}
\end{equation}
\begin{equation}
u_3(x,t)=-2k^{4},
\label{eq6}
\end{equation}
\begin{equation}
u_4(x,t)=-32k^{4},
\label{eq7}
\end{equation}
\begin{equation}
u_5(x,t)=-32k^{4}.
\label{eq8}
\end{equation}

It is wonderful but even trivial "solutions" \eqref{eq4}, \eqref{eq6}, \eqref{eq7} and \eqref{eq8} do not satisfy Eq.\eqref{eq3} at $C=0$.

Inan and Ugurlu in \cite{Ugurlu} also presented the exact solutions of the modified Burgers equation
\begin{equation}
u_{t}+u^{2}\,u_{x}+u_{xx}=0
\label{eq9}
\end{equation}

Authors \cite{Ugurlu} have used again the traveling wave solutions for Eq. \eqref{eq9}. They also have omitted the constant of integration in reduction of \eqref{eq9}. Finally, Inan and Ugurlu have considered following equation
\begin{equation}
w\,u+\frac{k}{3}\,u^{3}+k^{2}u'=0\,.
\label{eq10}
\end{equation}
Equation \eqref{eq10}  is well known and was introduced by Jakob Bernoulli in 1695. Eq. \eqref{eq10} can be reduced to the linear form using transformation
\begin{equation}
u(\xi)=\frac{1}{\sqrt{(v(\xi))}}
\label{eq10a}
\end{equation}

In this case Eq.\eqref{eq10} is reduced to the form
\begin{equation}
k^2\,v_{\xi}=\frac{2\,k}{3}+2\,w\,v
\label{eq10b}
\end{equation}

The general solution of Eq.\eqref{eq10} can be presented in the form
\begin{equation}
u(\xi)=\pm\,{\frac {3}{\sqrt {9\,{C_2}\,{{\rm e}^{{\frac {2\,w\xi}{{k}^{2}}
}}}-\,{\frac {3\,k}{w}}}}}
,
\label{eq10a}
\end{equation}
where $C_2$ is a arbitrary constant.

In \cite{Ugurlu} Inan and Ugurlu applied again the Exp-function method and found "three solutions" of Eq. \eqref{eq9}. The first "solution" is
\begin{equation}
u_1(x,t)=\frac{\sqrt{\frac{3k}{2}} \exp(kx-\frac{k^{2}}{2}t)-\sqrt{\frac{3k}{2}}b_{0}}{\exp(kx-\frac{k^{2}}{2}t)+b_{0}}
\label{eq11}
\end{equation}
However, solution \eqref{eq11} do not satisfy Eq. \eqref{eq9}. Substituting function \eqref{eq11} into Eq. \eqref{eq9} we obtain the following expression
\begin{equation}
E=-\frac {2 \sqrt {6}\, b_{0}^{2}\,k^{5/2} \exp(kx-\frac{k^{2}}{2}t ) \left( 2 \exp(kx-\frac{k^{2}}{2}t )- b_{0} \right) }{ \left( \exp(kx-\frac{k^{2}}{2}t )+ b_{0} \right) ^{4}}
 \label{eq12}
\end{equation}
We can see that this expression is not equal to zero in the general case.

The second and the third solutions can be written as
\begin{equation}
u_2(x,t)=\frac{\sqrt{3k} \exp(kx-\frac{k^{2}}{2}t)}{\exp(kx-\frac{k^{2}}{2}t)}
\label{eq13a}
\end{equation}
\begin{equation}
\begin{gathered}
u_3(x,t)=\frac{\sqrt{\frac{3k}{2}} \exp(kx-\frac{k^{2}}{2}t)+\sqrt{\frac{3k}{2}}b_{0}+\sqrt{\frac{3k}{2}}\,b_{-1}\exp(-kx+\frac{k^{2}}{2}t)}
{\exp(kx-\frac{k^{2}}{2}t)+b_{0}+b_{-1}\exp(-kx+\frac{k^{2}}{2}t)}
\end{gathered}
\label{eq14a}
\end{equation}
However these expressions can be presented as the trivial solutions of Eq.\eqref{eq10}. We have
\begin{equation}
u_2(x,t)=\sqrt{3k},
\label{eq13}
\end{equation}
\begin{equation}
\begin{gathered}
u_3(x,t)=\sqrt{\frac{3k}{2}}.
\end{gathered}
\label{eq14}
\end{equation}

However the trivial solutions \eqref{eq13} and \eqref{eq14} do not satisfy Eq.\eqref{eq10} as well.

We can see that all exact solutions obtained by Inan and Ugurlu in \cite{Ugurlu} are trivial and we can obtain these solutions without the Exp-function method. What is more these trivial solution do not satisfy equations as well.

In conclusion we strongly recommend to authors and referees to look at papers \cite{Kudryashov09a, Kudryashov09b, Kudryashov09c, Kudryashov09d, Kudryashov09e, Kudryashov10a, Kudryashov10b, Kudryashov10c, Kudryashov10d, Kudryashov10aa, Kudryashov10bb, Kudryashov10e, Parkes09a, Parkes09b, Parkes09c, Parkes10a, Parkes10b, Popovych, Disine} carefully before finding exact solutions of nonlinear differential equations.

\end{document}